\documentclass[prb,aps,epsf]{revtex4}
 
\begin{document}

\title{Self compensation of classical non abelian charge}

\author{E.A. Bartnik\thanks{bartnik@fuw.edu.pl}}

\address{Institute of Theoretical Physics, Warsaw University, Ho\.{z}a Street 69, 00-681 Warsaw, Poland}

\begin{abstract}
A new classical, non singular solution with arbitrarily low energy is found 
for SU(2) non abelian fields
in the presence of a  static charge. Physically it means that a classical charge coupled to any SU(N)
non abelian gauge field will develop a pure gauge field, carrying no energy, that will
completely screen it - there are no visible classical non abelian charges.

\end{abstract}

\pacs{field theory, QCD, gauge fields}

\maketitle

\section{Introduction}

It is known that a classical (non quantal) system consisting of a static charge coupled
to an non abelian gauge field has a solution which is identical to electrodynamics
- the charge develops an electric field according to (abelian) Gauss Law \cite{Bogo}. 
By suitable Lorenz and gauge transformations one obtains easily fields for a moving charge,
in arbitrary gauge.
However, due to non linear coupling of non abelian SU(N) fields another solutions exists: 
the charge 
develops a pure gauge field which screens it
\cite{sc78}
\cite{sc81}
\cite{sc91}
 This configuration has energy arbitrarily close 
to zero. This means that in the classical gauge theory there can be no visible 
(i.e. interacting) charges. We explicitly construct a  new self screened solution for the case of 
a given static charge distribution coupled to an SU(2) gauge field. Since SU(2) is a subgroup
of any SU(N) group, the result is quite general. 

\section{Explicit Solution}

The Lagrange density of an non abelian gauge field  $A^{a}_{\mu}$ coupled to an conserved 
current   $j^{a}_{\mu}$  (we use Einstein summation convention throughout) is

         \begin{equation}
 {\cal{L}}=-\frac{1}{16\pi }F^{\mu\nu}_{a} F_{\mu\nu}^{a} -j^{\mu}_{a} A_{\mu}^{a}
 ,
        \end{equation}

where

        \begin{equation}
F_{\mu\nu}^{a}=\partial_\mu A_{\nu}^{a}-\partial_\nu A_{\mu}^{a}
               +g f^{a b c} A_{\mu}^{b} A_{\nu}^{c}
        \end{equation}
        
$f^{a b c}$  are SU(N) structure constants and $a,b,c= 1,2....(N^{2}-1)$ are group indices.
In the case of SU(2) $f^{a b c}$  is the totally antisymmetric symbol
${\varepsilon}^{a b c}$ in 3 dimensions.\

A suitable choice of gauge can simplify the problem significantly. In the case of static
charges it is judicious to choose the class of temporal gauges

        \begin{equation}
A_{0}^{a}=\varphi^{a}=0
        \end{equation}

and use non relativistic (3 dimensional) notation. Using electrical fields 

       \begin{equation}
\vec{E}^{a}=\frac{\partial \vec{A}^{a}}{\partial t} 
        \end{equation}
        
and magnetic fields

        \begin{equation}
\vec{B}^{a}=\vec{\nabla} \times \vec{A}^{a}+ g f^{a b c} \vec{A}^{b} \times \vec{A}^{c}         
        \end{equation} 
        
the Lagrange density takes a simple form

        \begin{equation}
{\cal{L}}=\frac {\vec{E}_{a} \cdot \vec{E}^{a}}{2}-\frac {\vec{B}_{a} \cdot \vec{B}^{a}}{2}
 +\vec{j}_{a} \cdot \vec{A}^{a},
        \end{equation}
        
Notice that in this gauge canonical variables are vector fields $\vec{A}^{a}$ and that
their canonical momenta are simply the electrical fields  $\vec{E}^{a}$. However not all 
configurations are allowed: varying  Lagrange density eq. 1 with respect to $\vec{A}^{a}$ 
one finds that fields have to fulfill the (non abelian) Gauss Law

\begin{equation}
\frac{\delta \cal{L}}{\delta A_0^{a}} = 0 \Rightarrow  \,\,\,\,\,
\vec{\nabla} \vec{E}^{a} + g f^{a b c} \vec{A}^{b} \cdot \vec{E}^{c} = 4 \pi \rho^{a} 
\end{equation}.

where $\rho^{a}$ is the charge density.

Cannonical Hamiltonian density is

        \begin{equation}
{\cal{H}}=\frac {\vec{E}^{a} \cdot \vec{E}^{a}}{2}+\frac {\vec{B}^{a} \cdot \vec{B}^{a}}{2}
 -\vec{j}_{a} \cdot \vec{A}^{a},
        \end{equation}.
        
Within this scheme the problem of fields in the presence on an external charge distribiution
can be formulated as follows: find fields $\vec{A}^{a}$, $\vec{E}^{a}$ minimalising
the energy

\begin{equation}
E= \int d^{3}r \; {\cal{H}}
\end{equation}

while fulfilling exactly Gauss Law eq. 7. Next one has to solve the equations of motion
with the above mentioned configuration as initial condition.
\\
For definitness let us assume we have a regular (i.e. non singular) radial charge
distribution in direction 1 in group space

\begin{equation}
\varrho^{a}(r)=
\left(
\begin{array}{cc}
\varrho(r) \\
0 \\
0 
\end{array} 
\right)
\end{equation}

We first perform a global time independent gauge transformation
- a small rotation around direction 2. Now charge density is

\begin{equation}
\varrho^{a}(r)=
\left(
\begin{array}{cc}
c \varrho(r) \\
0 \\
s \varrho(r)
\end{array} 
\right)
\end{equation}
        
where $c=\cos(\alpha), s=\sin(\alpha)$ with $\alpha \ll 1$.    
Choose $\vec{A}^{a}$ to be pure gauge - a spatial gradient of a function
along direction 2 in group space

 \begin{equation}
\vec{A}_{a}=
\left(
\begin{array}{cc}
 0\\
\vec{A}_2 \\
0
\end{array} 
\right)
=
\left(
\begin{array}{cc}
 0\\
\nabla \chi(r) \\
0
\end{array} 
\right)
\end{equation}  

Gauss Law eq.7 simplifies to

\begin{eqnarray}
 \nabla \vec{E}_1+ g \vec{A}_2 \cdot \vec{E}_3 = 4 \pi c \varrho (r)  \nonumber\\
 \nabla \vec{E}_2=0            \\
 \nabla \vec{E}_3- g \vec{A}_2 \cdot \vec{E}_1 = 4 \pi s \varrho (r)   \nonumber
\end{eqnarray}

It is obvious that $ \vec{E}_2 = 0$. Now we choose $ \vec{E}_1 = 0$ so eq.'s (13) simplify to

 \begin{eqnarray}
  g \vec{A}_2 \cdot \vec{E}_3 = 4 \pi c \varrho (r)  \\
 \nabla \vec{E}_3 = 4 \pi s \varrho (r)   
\end{eqnarray}

Equation 15 is used to determine $\vec{E}_3$ while eq.14 gives $\vec{A}_2 $. 
Since $\vec{E}_3$ is generated (in total analogy to electrodynamics)
by charge distribution $\varrho$ multiplied by arbitrarily small
parameter s, its contribution to the energy of the system is (quadratically in s) small. Equation 
14 determines pure gauge field $\vec{A}_2 $ which screens the overwhelming majority of charge
$\varrho$ without giving any contribution to the energy of the system. Thus we have found 
an initial field configuration which satisfies exactly Gauss Law eq.7 and has an arbitrarily
small energy - vector potential $\vec{A}_2$ screens most of the charge. To be very explicit let
us write the solution of eq.15

\begin{equation}
\vec{E}=(\frac{\vec r}{r})  \frac{ s Q(r)}{r^2}
\end{equation}

where $Q(r)$ is the charge contained in the sphere of radius $r$

\begin{equation}
Q(r)=4 \pi \int_{0}^{r} s^2 \varrho (s) ds
\end{equation}

Pure gauge field $\vec{A}_2 $ is readily obtained with help of eq. 14

\begin{equation}
\vec{A}_2=\nabla \chi (r)= (\frac{\vec r}{r}) \chi\prime(r)
\end{equation}

where

\begin{equation}
 \chi\prime(r) = (\frac{c}{s g}) \frac {r^2 \varrho(r)}{Q(r)}
               = (\frac{c}{s g}) \frac {Q\prime(r)}{Q(r)} 
\end{equation}

It follows that the pure gauge compensating field $\vec{A}_2 $ exists only in regions where 
charge density is non vanishing. Assuming that the charge distribution $\varrho(r)$ is regular 
at origin

\begin{equation}
\varrho(r) \approx \varrho_{0} \hspace{5mm} r\ll 1
\end{equation}

we find

\begin{equation}
\vec{A}_2= (\frac{\vec r}{r}) \chi\prime(r) \approx (\frac{\vec r}{r}) \frac{3}{r}
\end{equation}

which is nonsingular due to $r^2 dr$ element of radial integration. Notice also that 
$\chi$ is independent of the value of charge density $\varrho_0$.
In fact we have quite generally

\begin{equation}
\chi(r)=(\frac{c}{s g}) \ln (Q(r)).
\end{equation}

The Hamilton equations of motion for canonically conjugated 
fields $\vec{A}^{a}$ and $\vec{E}^{a}$,

       \begin{equation}
\frac{\partial \vec{A}^{a}}{\partial t}=\vec{E}^{a}, 
        \end{equation}

      \begin{equation}
\frac{\partial \vec{E}^{a}}{\partial t}=-\nabla \times \vec{B}^{a}
                                        -2 g f^{a b c} \vec{B}^{b} \times \vec{A}^{c} 
        \end{equation}
        
are simply solved:

 \begin{equation}
\vec{E}_{a}=
\left(
\begin{array}{cc}
 0\\
 0\\
\vec{E}_3
\end{array} 
\right)
=
\left(
\begin{array}{cc}
 0\\
 0\\
(\frac{\vec r}{r})  \frac{ s Q(r)}{r^2}
\end{array} 
\right)
\end{equation}

i.e. electic fields are constant when magnetic fields eq.5 are zero; vector potentials are
simply

 \begin{equation}
\vec{A}_{a}=
\left(
\begin{array}{cc}
 0\\
\vec{A}_2 \\
\vec{A}_3
\end{array} 
\right)
=
\left(
\begin{array}{cc}
 0\\
\nabla \chi(r) \\
t \vec{E}_3
\end{array} 
\right)
\end{equation}  

we easily see that those vector potentials produce no magnetic field.
\\
    For point charges $\varrho$ is a 3 dimensional Dirac delta distribution

\begin{equation}
\varrho(r)=Q_{0}\,\, \delta_{3}(\vec{r}) 
\end{equation}

which can be understood as a point limit of a Gaussian charge distributon

\begin{equation}
\varrho(r)=Q_{0}\,\, \lim_{\sigma \rightarrow 0}
{(2 \pi \sigma)^{3/2} \exp(-\frac{\vec r^{2}}{\sigma^{2}})}
\end{equation}

Now $\chi$ is also a distribution with support at point $\vec r=0$ 
and is given by an analogous point limit of eq. 22.

 \section{Conclusions}
 
     We have found an explicit solution of SU(2) Yang-Mills field theory in the presence 
 of classical external charge. The fields screen the charge resulting in a configuration
 consisting of a charge and a pure gauge field. Such a configuration is static in the sense
 that the energy carrying electric fields $\vec E^{a}$ are time independent (vector potentials
 $\vec A^{a}$ rise linearly with time) and carries negligible energy. 
 \\
 At this stage we can only speculate on the situation in QCD.
 While it is obvious that the color charge of a quark (belonging to {\bf3} representation) 
 cannot be neutralized by any amount of gluons (members of {\bf8} representation) it is conceivable
 that large color charges would be screened - large color charge separation in high energy
 nucleus-nucleus collisions would be an impossible goal.


\begin{thebibliography}{99}
 \bibitem{Bogo} P.N. Bogolyubov, A.E. Dorokhov,Theor. and Math. Phys, V51 (1982) 462-468
 \bibitem{sc78} P. Sikivie, N. Weiss, Phys. Rev. Lett. 40 (1978), 1411-1413
 \bibitem{sc81} Rosy Teh et al. , Phys. Rev. D23, (1981) 3046-3049
 \bibitem{sc91} W.B. Campbell, R.E. Norton, Phys. Rev. D44 (1991) 3931-3934
 \end{thebibliography}
\end{document}